\documentclass{elsart}
\usepackage{graphicx}
\usepackage{amsmath}
\usepackage{amssymb}
\newcommand{\GeVcc}    {\mbox{$ {\mathrm{GeV}}/c^2                           $}}

\newcommand{\hetrois}    {\mbox{$ ^{3}{\mathrm{He}}                            $}~}
\newcommand{\hetro}    {\mbox{$ ^{3}{\mathrm{He}}                            $}}

\newcommand{\hequatre}    {\mbox{$ ^{4}{\mathrm{He}}                            $}~}

\newcommand{\neut}{$\tilde{\chi}$~}
\newcommand{\neutt}{$\tilde{\chi}$}

\newcommand{\gam} {{$\gamma$-ray}~} 
\newcommand{\gams} {{$\gamma$-rays}~}

\def\NIMA#1#2#3{{\rm Nucl.~Instr.~and~Meth.} {\bf{A#1}} (#2) #3}

\def\PLB{{\em Phys. Lett.}  B}
\def\PRD#1#2#3{{\rm Phys. Rev.} {\bf{D#1}} (#2) #3}
\def\PRB#1#2#3{{\rm Phys. Rev.} {\bf{B#1}} (#2) #3}

\def\PRL#1#2#3{{\rm Phys.~Rev.~Lett.} {\bf{#1}} (#2) #3}
\def\PLB#1#2#3{{\rm Phys.~Lett.} {\bf{B#1}} (#2) #3}

\def\APJ#1#2#3{{\rm Astrophys.~J.} {\bf{#1}} (#2) #3}

\begin{document}
\begin{frontmatter}
\title{Search for Supersymmetric Dark Matter with Superfluid \hetrois (MACHe3)}
\author[isn]{F. Mayet\thanksref{corr}},
\author[isn]{D. Santos},
\author[crtbt]{Yu. M. Bunkov},
 \author[crtbt]{E. Collin},
\author[crtbt]{H. Godfrin}
\thanks[corr]{Corresponding author : Frederic.Mayet@cea.fr (phone: +33 1-69-08-45-55, fax:+33 1-69-08-64-28)}
\address[isn]{Institut des Sciences Nucl\'eaires, 
 CNRS/IN2P3 and Universit\'e Joseph Fourier, 
 53, avenue des Martyrs, 38026 Grenoble cedex, France}
\address[crtbt]{Centre de Recherches sur les Tr\`es Basses Temp\'eratures,
 CNRS and Universit\'e Joseph Fourier, BP166, 38042 Grenoble cedex 9, France}  
\begin{abstract}
MACHe3 (MAtrix of Cells of superfluid \hetro) is a project of a new detector for direct Dark Matter search, 
using superfluid \hetrois as a sensitive medium. This paper presents a phenomenological study done 
with the DarkSUSY code, in order to investigate the discovery potential of this project of detector, as well as its
complementarity with existing and planned devices.
\end{abstract}
\end{frontmatter}
%
%
\section{Introduction}
A substantial body of astrophysical
evidence supports the existence of non-baryonic Dark Matter (DM) in the halo of our galaxy, in particular in the form of new, yet
undiscovered, weakly interactive massive particles (WIMPs)~\cite{jungmanphysrep}. One of the leading 
candidates is the neutralino predicted by the supersymmetric
extensions of the Standard Model of particle physics.\\
Following early experimental works~\cite{lanc}, a superfluid \hetrois detector has 
recently been proposed \cite{firstmac3} for direct Dark Matter search. Monte Carlo simulations have 
shown that a high granularity detector, a matrix of superfluid \hetrois cells, would allow to reach a high rejection factor
against background events, leading to a low false event rate. The purpose of this paper is to present a full 
estimation of the neutralino (\neutt) event rate, within the framework of the Minimal Supersymmetric Standard Model (MSSM), in order to compare 
with background event rate obtained by Monte Carlo simulation~\cite{firstmac3,daniel}. Finally, the complementarity
with existing devices, both for direct and indirect detection, will be shown.

\subsection{Experimental device}
The elementary component of MACHe3 is the superfluid \hetrois cell~\cite{prl95,prb98}. It is a small copper cubic box 
($\mathrm{V} \simeq 125 \,{\rm mm}^3$) filled with superfluid \hetrois in the B-phase. This ultra low temperature device (T $\simeq \!100 \,\mu\mathrm{K}$) presents a low detection 
threshold ($\mathrm{E}_{th}\simeq 1 \,\mathrm{keV}$). An experimental test of such a prototype cell has been done at 
CRTBT in June 2001.
Preliminary results \cite{thesefmayet,santosgamma} show that a threshold value down to $\sim 1 \,{\rm keV}$ 
has been achieved, and that a stability of the order of one week, at T $\simeq \!100 \,\mu\mathrm{K}$, has been obtained.\\
The final version of the detector will be a matrix of 1000 cells of $125 \,{\rm cm}^3$ each. The idea is to take advantage both on the 
energy loss measurement and the correlation among the cells to discriminate neutralino events from those of background 
(neutrons, \gams and muons). The design of the matrix has been optimized, with a Monte Carlo 
simulation \cite{firstmac3}. 
In the preferred configuration, a $10 \,{\rm kg}$ detector, the false event rate has been shown to be as small as 
$\sim\! 10^{-1}\, {\rm day}^{-1}$ for neutron events\footnote{Fast neutron contribution, from interaction of muons 
in the rock, is expected to be negligible as the enregy release in the cell will be much greater than $6\,{\rm kev}$.} and $\sim\! 10^{-2}\, {\rm day}^{-1}$ for muon events 
\cite{firstmac3}. Background from \gam events needs to be 
taken into account. Energy loss measurement and correlation among the cells within a $10 \,{\rm kg}$ detector 
allows to obtain a rejection up to 99.8\% for $\sim 2\,{\rm MeV}$ $\gamma$-rays \cite{firstmac3}. Additionnal internal tag on $\gamma$-rays may be obtained using 
a new matrix configuration 
in which two neigbouring cells share a common copper wall, thus greatly improving correlation factor while reducing the amount
of copper used. Monte Carlo studies are under way \cite{santosgamma}. Furthermore, $\gamma$-ray contamination 
is to be estimated in the forthcoming months for a multicellular prototype matrix.\\
In order to compare these false event rates with the expected neutralino rate, only muons and neutrons will be 
taken into account in the following. We shall recall that neutrons are usually considered as the ultimate background noise for 
this type of search, as they interact {\it a priori} like WIMPs.

\subsection{\hetrois as a sensitive medium for direct DM search}
Several properties of \hetrois make this nucleus a promising candidate for a sensitive medium for 
direct DM search.

{\bf a)} Concerning background rejection, and as outlined in \cite{firstmac3,york}, the neutron capture process offers the
possibility to discriminate neutron and \neut event, when considering a ${\rm 10\,
kg}$ granular detector. Compton cross-section being small ($\sigma \lesssim 1\,{\rm barn}$), the interactions with \gams will be minimized.
Eventually, as explained in \cite{prb98}, superfluid \hetrois is produced with an extremely high purity, the only 
solute being \hequatre in a negligible fraction. Consequently, no contamination  from radioactive materials is expected in the
sensitive medium. Of course, natural radioactivity from external materials (${\rm Cu}$, ...) has to be taken into account, by a
careful selection of these materials.

{\bf b)} Concerning neutralino detection the advantage is twofold. First, the maximum recoil energy does only slightly depend on the
neutralino mass, due to the fact that the target nucleus (${\rm m=2.81 \, GeV}\!/c^2$) is much
lighter than the incoming  \neut (${\rm M_\chi \geq 32 \,GeV}\!/c^2$), considering latest results from collider 
experiments \cite{pdg2000}. As a matter of fact, the recoil
energy range needs to be studied only below ${\rm 6 \,keV}$, see \cite{firstmac3,thesefmayet}. 
Second, \hetrois being a 1/2 spin nucleus, an \hetrois detector will be sensitive mainly 
to axial interaction, making this device complementary to existing ones, as shown below. In fact, the axial
interaction is largely dominant (up to three orders of magnitude) in all the SUSY region associated 
with a substantial elastic cross-section \cite{thesefmayet}.

%
%

\subsection{Theoretical framework}
\label{sec:theo} 
This phenomenological study has been done with the DarkSUSY code\footnote{The version used is 3.14.01,
with correction of some minor bugs.} \cite{ds}, within the framework of the 
phenomenological Supersymmetric Model, namely with the following free parameters : 
\begin{equation}
 \mu ,\; \mathrm{M}_2 ,\; \tan \beta ,\; m_{A} ,\; m_{0} \;\mathrm{and} \; A_{b,t}
\end{equation}

\noindent
with $\mu$ ($\mathrm{M}_2$) the Higgsino (Gaugino)  mass parameter, $\tan \beta$ the ratio of Higgs vacuum expectation values, 
$m_{A}$ the CP-odd Higgs boson mass, $m_{0}$ the common scalar mass and $A_{b,t}$ the soft trilinear coupling 
parameters\footnote{For a good introduction to MSSM models, we refer the reader to \cite{djouadi}.}.\\
Apart from $A_{b,t}$ chosen at fixed zero value, as their influence is expected to be negligible,  all the parameters have been scanned on a large range, with a variable
number of steps (tab.~\ref{tab:scan}). This scan of the free parameters corresponds to a total number of supersymmetric (SUSY) 
models of the order of $2 \times 10^6$.\\ 
We shall suppose all through this work the  
neutralino (\neutt), the lightest supersymmetric
particle, as the particle making up the bulk of galactic cold DM.
Each SUSY model is then checked not to be excluded by collider experiments~\cite{pdg2000}, including $b \to s\gamma$ limit.

The nest step, the evaluation of the relic density, is the key point of any Dark Matter calculations. Given the number of 
free parameters (5 in our case), the allowed SUSY parameter space may be extremely large, leading to 
a \neut relic density ranging on up to five orders of magnitude \cite{thesefmayet}. In order to exclude SUSY models giving a \neut relic density too 
far away from  the estimated matter density in the Universe \cite{omegam} ($\Omega_{\mathrm{M}} \simeq 0.3$), only 
models with $\Omega_\chi$ in the following range are considered :
\begin{equation}
0.025 \leq \Omega_\chi \mathrm{h_0}^2 \leq 1
\label{eq:omega}
\end{equation}
\noindent where $\mathrm{h_0} = (0.71 \pm 0.07)\times^{1.15}_{0.95}$ is the normalized Hubble expansion 
rate~\cite{pdg2000}.\\
The lower limit comes from the condition that the neutralino relic density has to be at least 
greater that the baryonic density, and the upper limit is a  conservative limit so that \neut do 
not give a density greater than the Universe\footnote{It can be noticed that selecting on 
$\Omega_\chi \mathrm{h_0}^2$ allows for a slightly looser selection.}. 
The SUSY model is thus checked to really provide a good non-baryonic Dark Matter candidate. We follow
\cite{pbarberg} in the choice of the "cosmologically interesting" range of $\Omega_\chi$. It should be noticed that 
this loose selection allows to exclude a large number of models in our SUSY scan. 
For further details concerning the calculation of the \neut relic density, we refer the reader to \cite{ds}. 

As the detection on earth is concerned, a galactic halo model has to be considered. In the following, standard
parameters have been used, in a spherical isothermal halo distribution, 
with a local density ($\rho_0$) and an average velocity~($v_0$), with the following values :
\begin{equation}
\rho_0  \!=\!   0.3\,{\rm GeV}\!/c^2\,{\rm cm}^{-3}\, \;\mathrm{and}\;  \,v_0 \!=\! 220\,{\rm km}\,{\rm s}^{-1} 
\label{eq:theo}
\end{equation}  

\noindent
These parameters are widely used for dark matter detection computations, see \cite{rick} for instance.  
No clumpy galactic dark matter structures \cite{clumps} are considered hereafter, since the 
effect is expected to be small both for direct detection and neutrino telescopes, in which the signal depends on the 
local halo density, as emphasized in \cite{clumps2}.
\vspace*{-7mm}
\section{Spin dependent cross-section and event rate}
\label{sec:xsrate}
As early recognized by Goodman and Witten \cite{directold}, the interaction between neutralino and quarks may be 
either spin dependent or spin independent, involving different Feynman diagrams \cite{jungmanphysrep}. 
In the general WIMP case, the allowed interactions are : vectorial,
axial and scalar. The neutralino being a Majorana fermion, the vectorial interaction vanishes, leaving two
classes of interaction~: scalar and axial, the first one being
spin independent, the second spin dependent and obviously requiring  a non-zero spin nucleus. 
\hetrois being a light 1/2 spin nucleus, a medium made of such nuclei will be sensitive only to axial 
interaction, as shown in \cite{thesefmayet}.\\
Using the DarkSUSY code, the \neutt-\hetrois spin dependent cross-section has been evaluated. The calculation of the 
\neutt-quark elastic scattering amplitude is done at the tree-level, via an exchange of squark or $Z^0$. The amplitude on 
nucleon ($a_{p/n}$) is then evaluated by adding the contribution of each quark ($a_{qi}$), weighted by the quark contents 
of the nucleon : 
\begin{equation}
a_{p/n} = \sum_i  \cfrac {\Delta^{(p/n)}_i a_{qi}}{\sqrt{2} G_\mathrm{F}}
\end{equation}

\noindent
where $\Delta^{(p/n)}_i$ is the quark contents of the nucleon~\cite{adams}.\\
The axial cross-section on \hetrois is then given by :
\begin{equation}
\sigma_{spin}(^{3}{\mathrm{He}}) = \frac{32}{\pi} G^2_F m_r^2 \frac{(J+1)}{J} \left(a_p \!<\!S_p\!> + \;a_n  \!<\!S_n\!> \right)^2
\label{eq:xs} 
\end{equation}

\noindent
where $<\!S_{p/n}\!>$ is the spin contents of the \hetrois nucleus ($<\!S_p\!> = \!-0.05$ and $<\!S_n\!> = \!0.49$), 
$m_r$ is the reduced mass and $J$ the ground state angular momentum of the \hetrois nucleus.

\begin{figure}[p]
\begin{center}
\includegraphics[scale=0.8]{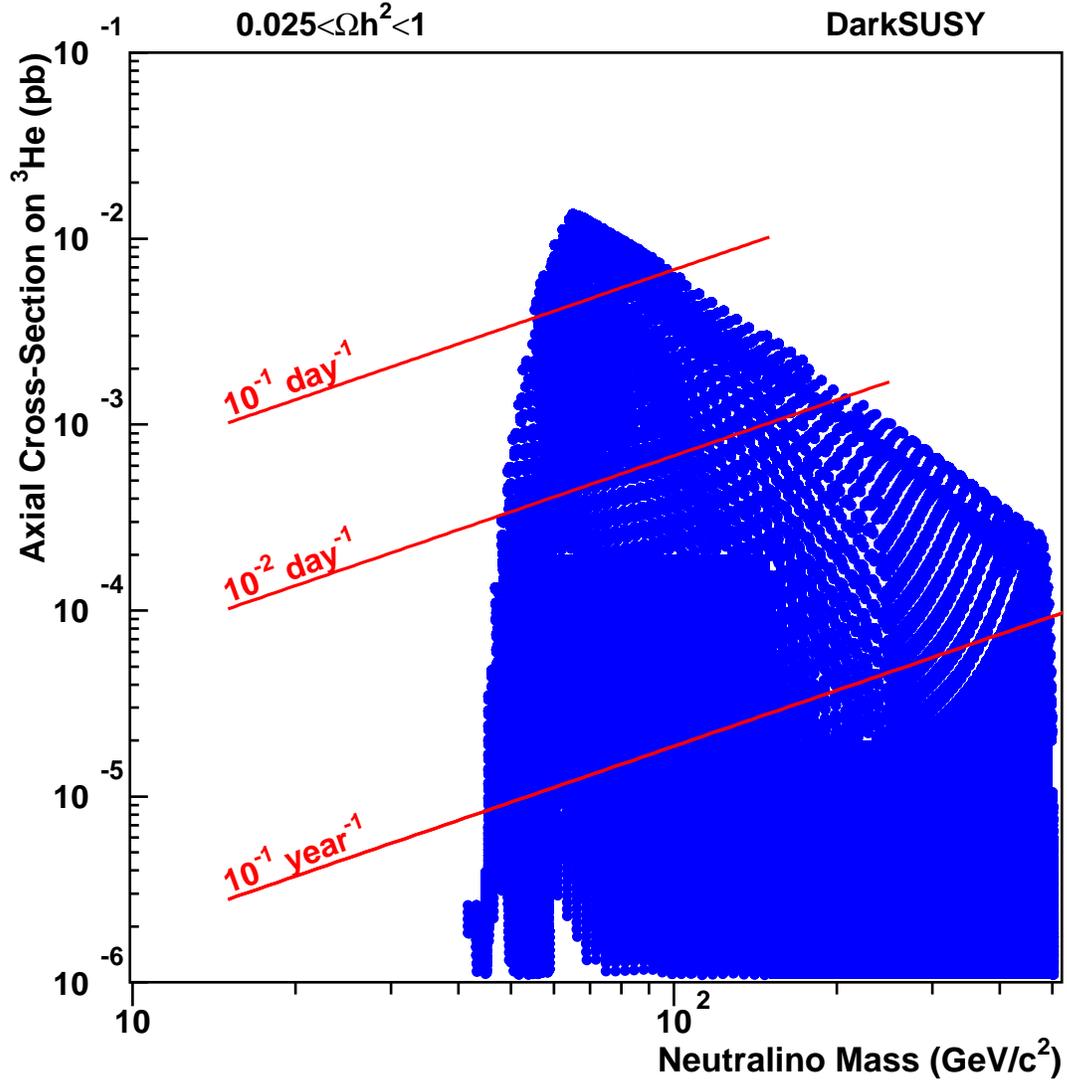}
\caption{Axial cross-section on \hetrois (${\rm pb}$) as a function of the neutralino mass (${\rm
Gev\!/}c^2$), for all models of the scan not excluded neither by colliders nor by 
cosmological constraints (eq.~\ref{eq:omega}). The rates corresponding to the Monte Carlo evaluation
\cite{firstmac3} of neutron and muon background are indicated for reference.}
\label{mac3:xs}
\end{center}
\end{figure}

Figure~\ref{mac3:xs} presents the cross-section on \hetrois as a function of the \neut mass. It can be seen that, 
for SUSY models not excluded by collider experiments and giving a relic density within the range of interest, a cross-section as
high as $\sim \!10^{-2} \,{\rm pb}$ can be obtained for $\sim \!60\,$\GeVcc~neutralino.\\
Using the cross-section, the \neut event rate (${\rm R}$) has been evaluated for a $10\,{\rm kg}$ \hetrois matrix, 
for comparison with the
background rate previously evaluated~\cite{firstmac3}.
\begin{equation}
{\rm R}\!=\! \cfrac{\rm \sigma(^{3}{\mathrm{He}})}{\rm M_{\chi}} \!\times \rho_{0}\!\times\! v_0 \!\times\!
\cfrac{\rm M_{det}}{\rm M_{He}}
\end{equation}
\noindent
with $\rho_0$ the local halo density and  $v_0$ the average neutralino velocity (eq.~\ref{eq:theo}).\\
A large number of models are giving a rate higher than the estimated false event rate induced by neutrons ( 
$\sim 0.1\;\mathrm{day}^{-1}$), or above the estimated muon background ($\sim 10^{-2}\;\mathrm{day}^{-1}$).
It could thus be concluded that a high granularity \hetrois detector would present a sensitivity to a large part of
the SUSY region.

In the following, the $\mu$ background level ($10^{-2}\;\mathrm{day}^{-1}$) is considered to be the lowest reachable limit
for MACHe3 and it is thus taken as the reference value. Any model giving a rate greater than this value is considered
hereafter to be {\em visible}. This will be used for the comparison with other DM search strategies.

%
%

\section{Complementarity with existing devices}
We present a study of the discovery potential of various detection strategies that may be correlated with
MACHe3. It is indeed worth understanding whether planned projects would be sensitive to different SUSY regions. In each case, we
will take into account  background estimations, or projected limits, in order to investigate the different discovery potentials. 
Choosing conservative values will allow to study the phenomenology of various searches without bias. It should be noticed 
that the results are model-dependent, both SUSY and galactic halo ones. However, standard galactic parameters
have been considered and the large SUSY scan allows for an exhaustive study within phenomenological MSSM
model.

\subsection{Complementarity with scalar detection}
\begin{figure}[p]
\begin{center}
\includegraphics[scale=0.75,angle=0]{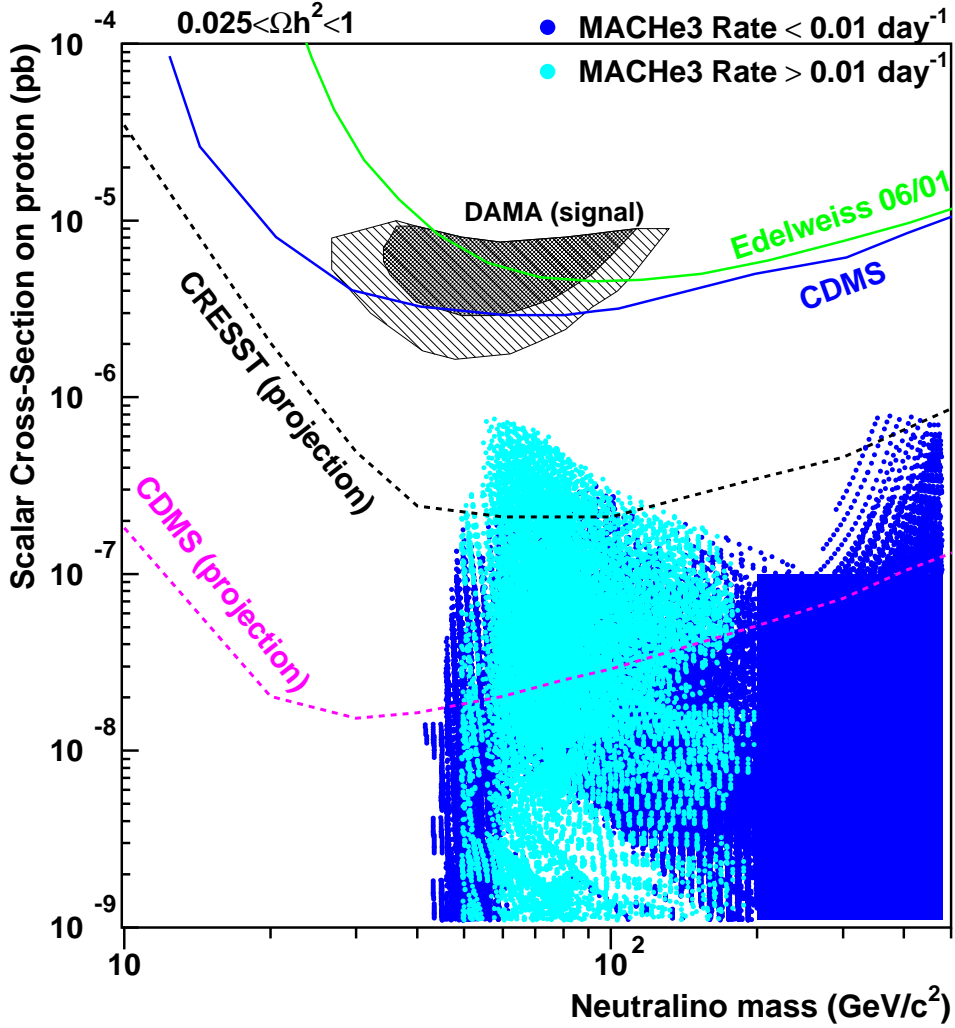}
\caption{Scalar cross-section (on proton) as a function of the \neut mass, for all models of the scan not 
excluded neither by colliders nor by cosmological constraints. Exclusion limits from the Edelweiss \cite{excledel} and 
CDMS \cite{exclcdms} experiments are shown, as well as the $3\,\sigma$ "DAMA region" \cite{damasig}. 
Dotted lines indicate projected limits \cite{rick} from
CRESST \cite{projcresst} and CDMS. 
Dark points indicate SUSY models giving a \neut rate in MACHe3 lower than the estimated background level, while light points are 
giving a rate higher than $10^{-2}\,{\rm day}^{-1}$. The two regions are overlapping as the signal in MACHe3 is not correlated with 
scalar cross-section.}
\label{fig:damamac3}
\end{center}
\end{figure}

Firstly, we studied the possible correlation between the two types of direct detection : axial and scalar. In the latter case,
many detectors are already running, or planned to start measurements in the near future. At first sight, these two direct searches should be largely
independent as they involve different processes, in particular the Feynman diagrams 
at the tree level are different.\\
Within the framework of the study described above, the proton scalar cross-section has been 
evaluated at the tree level\footnote{Higher order diagrams, involving gluon loops \cite{jungmanphysrep}, are not taken into account in the current version of
DarkSUSY.}. Figure~\ref{fig:damamac3} presents the scalar cross-section on proton as a function of the \neut mass, for all
models of the scan not excluded by collider experiments and associated with a relic density within the range of interest
(eq.~\ref{eq:omega}). It can be seen that many SUSY models lie below the projected limits of future scalar detectors 
(CDMS \cite{exclcdms}, CRESST \cite{projcresst}), while giving an event rate above the value taken as the lowest reachable limit for MACHe3.

\begin{figure}[p]
\begin{center}
\includegraphics[scale=0.8,angle=0]{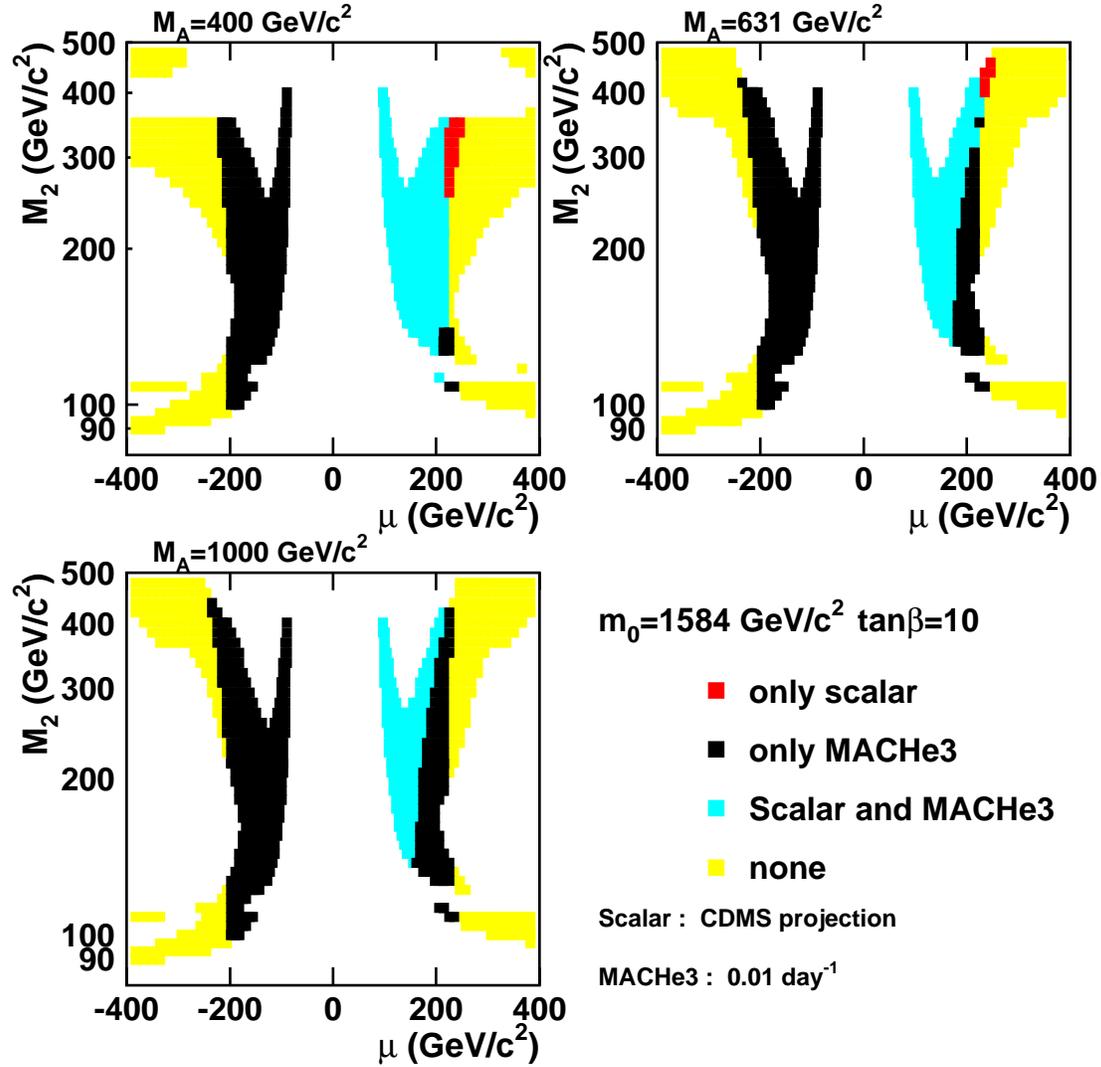}
\caption{Sensitivity regions for scalar search and MACHe3, in the (${\rm M_2, \mu}$) plane. The three
figures have been obtained with $\tan \beta = 10$, ${\rm m_0 \simeq \;1.6 \,TeV\!/}c^2$ and 
three values of ${\rm M_A}$. The central white region is excluded by collider limits, while the two
other white regions are associated with a relic density outside the preferred range. The color code
indicates the region potentially covered by MACHe3 (with a $10^{-2}\, {\rm day}^{-1}$ background
value), by scalar detection (for which CDMS projected limit is taken as ultimate limit), by both
strategies, or none of them.}
\label{fig:compdamamac3}
\end{center}
\end{figure}

As an illustration of this complementarity, we present on fig.~\ref{fig:compdamamac3} the regions potentially
covered by direct scalar search and by MACHe3, in a given part of the SUSY parameter space (see 
fig.~\ref{fig:compdamamac3} for details). The reference
limits have be chosen as : the CDMS projected limit \cite{rick} in one case and the $10^{-2}\,{\rm day}^{-1}$ limit for
MACHe3, previously discussed.\\ 
It can be noticed that only axial direct search may be sensitive to the $\mu\!<\!0$ region, in  this given 
part of the SUSY parameter space, due to the fact that scalar interaction vanishes. The cancellation arises in the  
$t$ channel (Higgs exchange) which is dominant, with u and d quark contributions of opposite 
sign and absolute values of the same order~\cite{sigspin}. Similar cancellation does arise 
for different value of $\tan \beta$, in the case of axial interaction. 
It can be concluded that these two detection strategies are sensitive, as expected, to
different SUSY regions, thus highlighting their complementarity.

\subsection{Complementarity with indirect detection}
Amongst various kind of indirect searches ($\bar{p}$, $\overline{\rm D}$, $e^+$, 
$\gamma$, $\nu$), the detection expected in neutrino
telescopes is the only one likely to be correlated with direct detection, as it involves both elastic scattering and
annihilation. The neutralino capture may occur either in the Sun or in the Earth, through successive elastic
scatterings. As far as comparison with direct detection is concerned, the key point is that the composition 
of the
Sun and the Earth are different. While the Earth is mainly made of even-even nuclei (O, Si, Mg, Fe), the 
Sun is
mainly made of H and He. So one expects the signal coming from annihilation in the centre of the Earth to be 
 only correlated with the signal expected from scalar direct detection. On the contrary, the Sun being 
made of both even and odd nuclei, a very light correlation is expected with direct detection, 
both scalar and axial.\\
Within the framework of the study described above, we tried to find out whether the MACHe3 project would present a
complementarity with the project of $\nu$ telescopes, such as IceCube \cite{icecube} or Antares \cite{antares}. 
These projects aimed at building a very large detector, of size of the order of $1\,{\rm km}^3$. The instrumentation
will thus be loose, giving a rather high detection threshold. We follow \cite{neutrinoberg} in choosing a reference 
threshold of $25\,{\rm GeV}$.

\begin{figure}[p]
\begin{center}
\includegraphics[scale=0.68,angle=0]{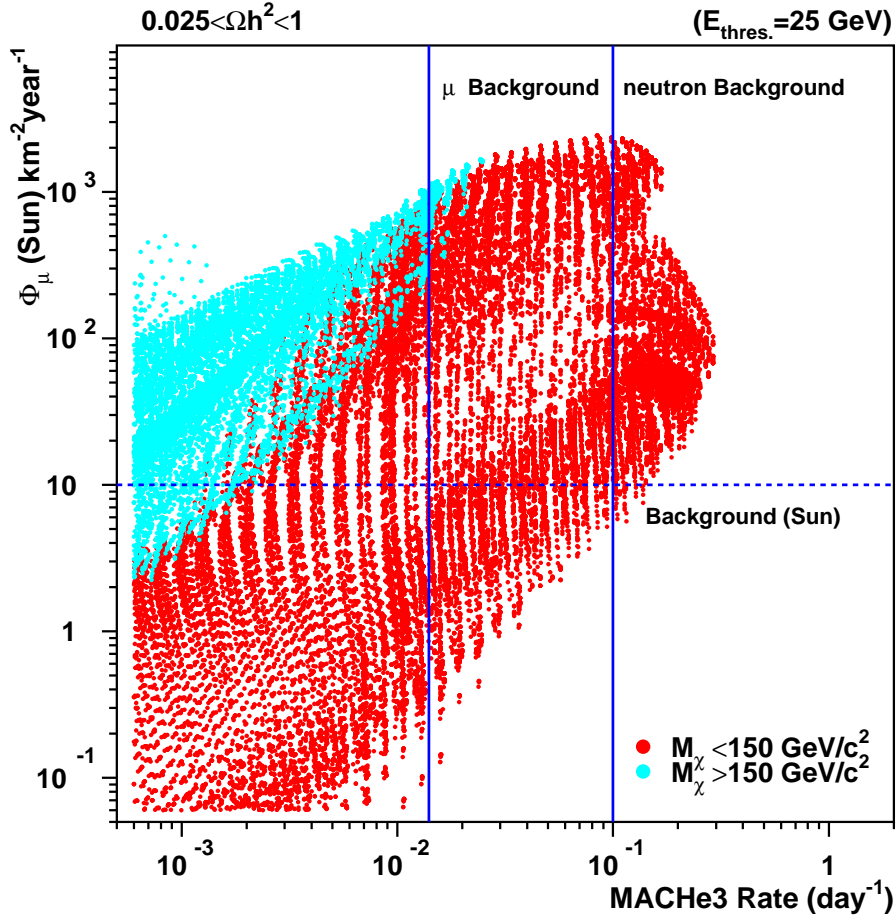}
\caption{Muon flux in neutrino telescopes (with a $25\,{\rm GeV}$ threshold) against event rate in MACHe3. For references, the
estimated backgrounds are shown, in one case from energetic neutrinos created in the sun's corona \cite{bdfnu}, and in the other case the
background from muon and neutrino events in MACHe3.}
\label{fig:neutrinosvsmache3}
\end{center}
\end{figure}

\begin{figure}[p]
\begin{center}
\includegraphics[scale=0.8,angle=0]{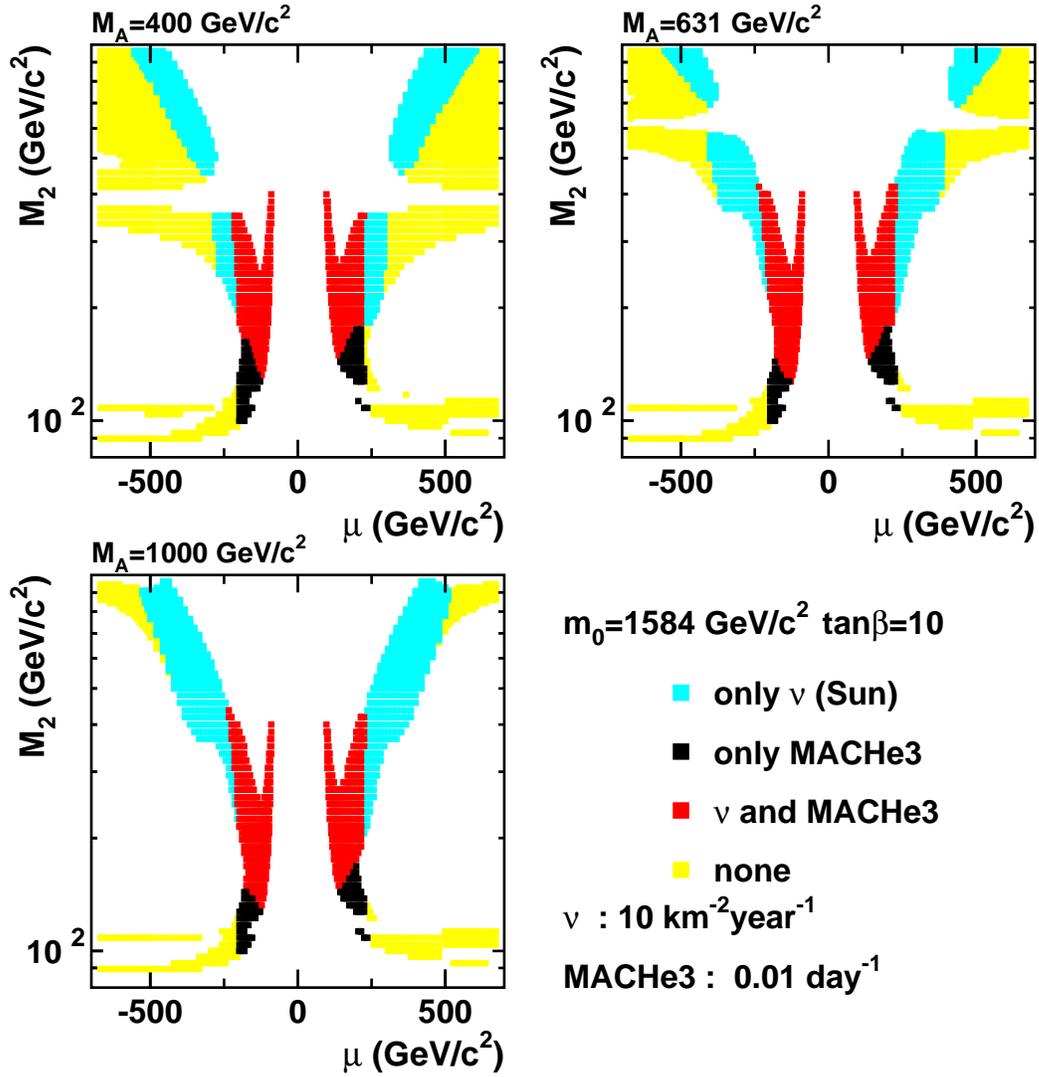}
\caption{See fig~\ref{fig:compdamamac3}. In this case the color code
indicates the region potentially covered by MACHe3 (with a $10^{-2}\, {\rm day}^{-1}$ background
value), by neutrino telescopes detection (for which a $10\,{\rm km^{-2}\,year^{-1}}$ background 
value \cite{bdfnu} is considered), by both strategies, or none of them.}
\label{fig:compantaresmac3}
\end{center}
\end{figure}

Figure~\ref{fig:neutrinosvsmache3} shows the result of the calculation for $\nu$ telescopes and MACHe3. The expected $\mu$
flux (${\rm km^{-2} \,year^{-1}}$) is shown against the rate in MACHe3 (${\rm day^{-1}}$). The background usually 
taken as ultimate for $\nu$ telescopes, coming from the interaction of cosmic rays in the sun's corona \cite{bdfnu}, is indicated at a level of 
$\sim \! 10\,{\rm km^{-2}\,year^{-1}}$. We consider this value as the most favourable limit for these projects of ${\rm km^{3}}$
detectors. In the same way, the background noise from $\mu$ in MACHe3 is taken as the lowest reachable level, as discussed
above (sec.~\ref{sec:xsrate}). Any model above
these limits are said hereafter to be {\em visible} for this type of detection. It can be concluded from 
fig.~\ref{fig:neutrinosvsmache3} that there is a complementarity between indirect $\nu$ detection and axial direct
detection, and in particular in the case of the MACHe3 detector. Many models may be {\em visible} in only one kind of detection 
strategy, high mass neutralinos being seen, for instance, only in $\nu$ telescopes.\\
As for the comparison between scalar and axial detection, we present on 
fig.~\ref{fig:compantaresmac3} the SUSY regions potentially observable for these two types 
of detection strategies. As expected, neutrino telescopes are sensitive to higher values of ${\rm M_2}$. 
For this given region of the SUSY parameter space, it can be concluded that the combination of these two
signals may close the gap between collider excluded region and cosmological bounded regions, 
up to high ${\rm M_2}$ values.

\section{Conclusion}
It has been shown that a ${\rm 10\,kg}$ high granularity \hetrois detector (MACHe3) would allow to obtain, in many SUSY models, a \neut event rate higher 
than the estimated (neutrons and muons) background. MACHe3 would thus 
potentially allow to reach a large part of the SUSY region, not excluded by current collider limits and for 
which the neutralino relic density lies within the range of interest. Furthermore, it has been shown that this project of new
detector would be sensitive to SUSY regions not covered by future or ongoing DM search detectors, both for direct and indirect
detection, thus highlighting the complementarity of MACHe3 with existing or planned devices.

\noindent \textbf{Acknowledgments : }\\
The authors wish to thank R.~Gaitskell and V.~Mandic for the convenient Dark Matter tools \cite{rick}.
One of us (FM) is grateful to G.~Coignet, Y.~Giraud-H\'eraud, L.~Mosca, G.~Sajot and C.~Tao, for 
fruitful discussions on this subject.

\listoffigures

\begin{table}[b]
\begin{center}
\begin{tabular}{|c|c|c|c|} \hline
Parameter       &       Minimum        & Maximum & Number of steps\\
\hline
 $\mid \!\mu\! \mid$ (GeV) &  50               &   1000 & 100 (+/-)\\
   $\mathrm{M}_2$ (GeV) &  50               &   1000  & 100\\
   $\mathrm{m}_0$ (GeV) &  100               &   $10^4$  & 11\\
   $\mathrm{M}_A$ (GeV) &  100               &   1000  & 3\\
 \hline
 $\tan \beta$  & \multicolumn{3}{|c|}{\raisebox{0pt}[12pt][6pt]{3, 10 and 60}}  \\
\hline 
\end{tabular} 
\caption{Scan of the SUSY parameters used for the study. The total number of models is of the order of $\sim 2 \times 10^6$.}
\label{tab:scan}
\end{center}
\end{table}

\newpage
\newpage

\newpage
\newpage

\begin{thebibliography}{99}
\bibitem{jungmanphysrep}G.~Jungman {\it et al.}, Phys.\ Rept.\  {\bf 267} (1996) 195
\bibitem{lanc}G.~R.~Pickett, in Proc. of the Europ. Workshop on Low Temp. Devices for 
the Detection of Low Energy Neutrinos and Dark Matter, 1988 (Annecy, France), Ed. L.~Gonzales-Mestres (Ed.
Frontieres) pp. 377
\bibitem{firstmac3}F.~Mayet {\it et al.}, \NIMA{455}{2000}{554}
\bibitem{daniel}D.~Santos {\it et al.}, Proc. of the 4th International Symposium 
on Sources and Detection of Dark Matter and Dark Energy in the Universe, 
February 2000, Marina Del Rey (CA, USA), Ed. D.~B. Cline, Springer, (astro-ph/005332)  
\bibitem{prl95}D. I. Bradley \textit{et al.}, \PRL{75}{1995}{1887} 
\bibitem{prb98}C. B\"{a}uerle \textit{et al.}, \PRB{57}{1998}{22}
\bibitem{thesefmayet}F.~Mayet, Ph. D. Thesis, Univ. of Grenoble (UJF, Grenoble, France), Sept. 2001
\bibitem{santosgamma}D.~Santos {\it et al.}, {\it in preparation}
\bibitem{york}F.~Mayet {\it et al.}, Proc. of the 3rd International  Workshop 
on the Identification of Dark Matter (IDM2000), Sept. 2000, York (UK),  Eds. N.~J.~C.~Spooner and
V.~Kudryavtsev, World Scientific 2001, (astro-ph/0011292)
\bibitem{pdg2000}D.~E.~Groom {\it et al.}, Eur.\ Phys.\ J.\ {\bf C15} (2000) 1, http://pdg.lbl.gov/
\bibitem{ds}P.~Gondolo {\it et al.}, {\it DarkSUSY}, in preparation,\\ http://www.physto.se/~edsjo/darksusy/download.html
\bibitem{djouadi}A.~Djouadi {\it et al.}, hep-ph/9901246
\bibitem{omegam}J.~R.~Primack, Nucl.\ Phys.\ Proc.\ Suppl.\  {\bf 87} (2000) 3
\bibitem{pbarberg}L.~Bergstr\"om {\it et al.}, \APJ{526}{1999}{215}
\bibitem{rick}R.~Gaitskell and V.~Mandic, http://dmtools.berkeley.edu/limitplots/
\bibitem{clumps}J.~Silk and  A.~Stebbins, \APJ{411}{1993}{439}
\bibitem{clumps2}L.~Bergstr\"om \textit{et al.}, Phys.\ Rev.\ {\bf D59} (1999) 043506
\bibitem{directold}M.~W.~Goodman and E.~Witten, \PRD{31}{1986}{3059}
\bibitem{adams}D.~Adams {\it et al.} (SMC Collaboration), \PLB{329}{994}{399}
\bibitem{excledel}M.~Chapellier {\it et al.}, Proc. of the 3rd International  Workshop 
on the Identification of Dark Matter (IDM2000), Sept. 2000, York (UK),  Eds. N.~J.~C.~Spooner and
V.~Kudryavtsev, World Scientific 2001, (astro-ph/0101204)
\bibitem{exclcdms}R.~Abusaidi \textit{et al.} (CDMS Collaboration), \PRL{84}{2000}{5699}
\bibitem{damasig}R.~Bernabei {\it et al.} (DAMA Collaboration), \PLB{480}{2000}{23}
\bibitem{projcresst}A.~Morales, Proc. of the International Workshop TAUP 99, Sept. 1999, Paris (France), Eds. J.~Dumarchez {\it et al.}, Nucl. Phys. {\bf B} 87 (2000) Proc. Suppl. 

\bibitem{sigspin}V.~A.~Bednyakov \textit{et al.}, \PRD{63}{2001}{095005}
\bibitem{neutrinoberg}L.~Bergstr\"om \textit{et al.}, \PRD{58}{1998}{103519}
\bibitem{icecube}F.~Halzen {\it et al.} (AMANDA Collaboration), in Proc. of the 26th International Cosmic Ray Conference, 
Aug. 1999, Salt Lake City (USA)
\bibitem{antares}R.~van Dantzig  (ANTARES Collaboration), Nucl.\ Phys.\ Proc.\ Suppl.\  {\bf 100} (2001) 341, and
references therein.
\bibitem{bdfnu}D.~Seckel \textit{et al.}, \APJ{382}{1991}{652}

\end{thebibliography}
\end{document}